\documentclass[preprint]{revtex4}

\usepackage{amsmath}
\usepackage{latexsym}
\usepackage{epsfig}
\usepackage{psfrag}
\usepackage{here}

\newcommand{\beq}{\begin{equation}}
\newcommand{\eeq}{\end{equation}}
\newcommand{\beqa}{\begin{eqnarray}}
\newcommand{\eeqa}{\end{eqnarray}}
\newcommand{\del}{\partial}

\newcommand{\vp}{``$+$''}
\newcommand{\vm}{``$-$''}

\begin{document}

%
%
%
%

\preprint{BI-TP 2005/47}

\title{Boltzmann Collision Term} 

\author{Steffen~Weinstock\footnote{e-mail: steffen@physik.uni-bielefeld.de}}
\affiliation{Fakult\"at f\"ur Physik, Universit\"at Bielefeld, 33615 Bielefeld, Germany\vspace{0.2cm}}



\begin{abstract}

We derive the Boltzmann equation for scalar fields using the Schwinger-Keldysh
formalism. The focus lies on the derivation of the collision term. We show that
the relevant self-energy diagrams have a factorization property. The collision
term assumes the Boltzmann-like form of scattering probability times
statistical factors for those self-energy diagrams which correspond to tree
level scattering processes. Our proof covers scattering processes with any
number of external particles, which come from self-energy diagrams with any
number of loops.

\end{abstract}

\maketitle

%
%
%
%

\section{Introduction}

%
%

Kinetic theory has proven to be a very successful tool for the description of
systems out of equilibrium.  Important applications can be found in many fields
of physics,
current examples are early cosmology, or the theory of heavy
ion collisions, where one aims to understand how the matter produced in the
collision evolves, and in particular whether it thermalizes or not.
A very famous kinetic equation is the Boltzmann equation for the evolution of
the particle distribution functions in a sufficiently dilute system.
A first derivation of non-relativistic kinetic theory from field theory was
given by Kadanoff and Baym~\cite{Kad62}.  Using the Closed-Time-Path formalism,
also called Schwinger-Keldysh formalism, this derivation becomes simpler and
can also be extended to relativistic theories, see for
example~\cite{Li:1982gk,Calzetta:1986cq,Mrowczynski:1992hq,Klevansky:1997wm,PSW}.
In early treatments the collision term of the Boltzmann equation was simply
modeled heuristically: consider each reaction the particle can undergo, compute
the probability for this reaction by using the free particle cross section, and
multiply with the appropriate statistical factors, that is the Bose-enhanced or
Fermi-suppressed phase space distribution functions, respectively.
The collision term for a scalar particle that can undergo 2-to-2 scattering,
for example, is
\beqa
   d_t f_p
 &=&
   \frac{1}{4\omega_p}
   \int
   \frac{d^3k}{(2\pi)^32\omega_k}
   \frac{d^3p'}{(2\pi)^32\omega_{p'}}
   \frac{d^3k'}{(2\pi)^32\omega_{k'}}
   (2\pi)^4\delta^4(p' +  k' - p - k)
\label{Boltzmann}
\\[1ex]&&\hphantom{XXX}
   \times
   {\cal P}_{pkp'k'}
   \times
   \big[ f_{p'} f_{k'} (1+f_p) (1+f_k)
        -(1+f_{p'}) (1+f_{k'}) f_p f_k
   \big]
.
\nonumber
\eeqa
We refer to a collision term of this form as a standard Boltzmann collision
term.
In
references~\cite{Li:1982gk,Calzetta:1986cq,Mrowczynski:1992hq,Klevansky:1997wm}
the collision term of the resulting Boltzmann equation was derived for a number
of specific theories from the right hand side of the Kadanoff-Baym equation
considering self-energy diagrams with up to two loops.
To our knowledge the first paper with a computation that went beyond the 2-loop
self-energy is~\cite{Carrington:2004tm}.  The authors explicitly computed a
selection of self-energy diagrams with up to three or four loops in real,
scalar $\phi^3$- and $\phi^4$-theory, respectively, and managed to bring them
into a form like~(\ref{Boltzmann}). This computation was quite involved and
required the assistance of a computer system to handle the big number of terms
appearing in intermediate steps.

%
%

In the present paper we consider a scalar theory with an unspecified
non-derivative self-interaction. We first show that the self-energies appearing
in the collision term have a useful factorization property.  In the strict
on-shell limit a certain class of self-energy diagrams then indeed leads to a
collision term of the form~(\ref{Boltzmann}), namely those that correspond to
tree level scattering diagrams with any number of external particles.
Not all contributions to the self-energy fit into this picture, which can
already be expected from the problems that arise when extending the vacuum
Cutkosky rules to finite temperature~\cite{Kobes,Gelis:1997zv}.

%
%

In section~\ref{sec:basics} we give the basics of nonequilibrium field theory
as far as required for our purpose. We also show briefly how the flow term of
the Boltzmann equation is obtained from the Schwinger-Dyson equation.
This procedure is standard and has extensively been covered in literature
already, so only the basic steps are given here. For more details see for
example~\cite{Carrington:2004tm} and references therein.
Section~\ref{sec:collision-term} is the main part of this paper, where we study
the collision term for a real scalar field.
A remark on the extension to charged scalar fields can be found in
section~\ref{sec:complex}, and in section~\ref{sec:discussion} we finally
discuss our results.

%
%
%
%

\section{Basics}
\label{sec:basics}

An appropriate framework to study the time evolution of physical quantities in
nonequilibrium situations is given by the Schwinger-Keldysh or Closed-Time-Path
formalism~\cite{Schwinger:1960qe,Keldysh:1964ud,Chou:1984es}. The basic
technical point is that the time variable of all objects is defined on a path
$C$ on the real axis that leads from $-\infty$ to $+\infty$, and then back to
$-\infty$. The definition of the Green function is still
\beq
  \Delta(x, y) = -i \left< T_C \phi(x) \phi^\dagger(y) \right>
,
\label{Green_function:def:path}
\eeq
but time ordering is performed along the path $C$ here. If we split the path
into a \vp-branch from $-\infty$ to $+\infty$ and a \vm-branch going back to
$-\infty$, we can distinguish four real time Green functions, differing by the
branches on which the time arguments are placed:
\beqa
      i\Delta^{++}(x, y)
  &=& i\Delta^t(x, y)
   =  \left< T \phi(x) \phi^\dagger(y) \right>
,
\label{Green_function:def:index}
\\
      i\Delta^{+-}(x, y)
  &=& i\Delta^<(x, y)
   =  \left< \phi^\dagger(y) \phi(x) \right>
,
\nonumber\\
      i\Delta^{-+}(x, y)
  &=& i\Delta^>(x, y)
   =  \left< \phi(x) \phi^\dagger(y) \right>
,
\nonumber\\
      i\Delta^{--}(x, y)
  &=& i\Delta^{\bar{t}}(x, y)
   =  \left< \bar{T} \phi(x) \phi^\dagger(y) \right>
 .
\nonumber
\eeqa
The $\bar{T}$ in the last line denotes anti time ordering.  These functions are
not independent of each other, but are connected via the relation
\beq
  \Delta^t + \Delta^{\bar{t}} = \Delta^< + \Delta^>
.
\label{Green_function:relation}
\eeq
In addition, one defines the retarded and advanced Green functions:
\beq
   \Delta^{R,A}(x, y)
 = \Delta^t(x, y) - \Delta^{<,>}(x, y)
.
\label{Green_function:def:ret_adv}
\eeq
In contrast to ordinary perturbation theory in vacuum, here each internal
vertex can be either of type \vp or of type \vm. The lines between vertices
represent the Green functions defined in~(\ref{Green_function:def:index}),
depending on the types of the incident vertices. For each vertex of type \vm
there is an additional factor $-1$.

%
%

The Schwinger-Dyson equation for the contour Green
function~(\ref{Green_function:def:path}) leads to the following equation of
motion:
\beq
    \left( \del^2_x + m^2 \right) \Delta(x,y)
  = - \delta^4_C(x-y)
    - \int_C d^4z \, \Pi(x,z) \Delta(z,y)
.
\label{Green_function:EOM:path}
\eeq
The self-energy $\Pi$ is defined as $i$ times the sum of all one-particle
irreducible two-point functions. We switch to index notation and obtain the
equation of motion for $\Delta^{<,>}$:
\beq
    \left( \del^2_x + m^2 \right) \Delta^{<,>}(x,y)
  = \int d^4z \,
     \left(\Pi^{<,>}(x,z) \Delta^A(z,y) - \Pi^R(x,z) \Delta^{<,>}(z,y) \right)
.
\label{Green_function:EOM:index}
\eeq
For simplicity we omit a potential tadpole contribution to the self-energy
$\Pi$. We assume that its only effect is a shift in the mass of the particles
which possibly introduces a force on the left hand side of the resulting
Boltzmann equation, but that doesn't change the argumentation concerning the
collision term.

%
%

Our goal is an equation that describes the evolution of the phase space density
of particles, which takes place on a macroscopical scale. In order to separate
this from the quantum evolution on a microscopical scale, we perform a Wigner
transformation. For any two-point function we introduce the average coordinate
\beq
  X = \frac 12 (x+y)
\label{averagw_coordinate:def}
\eeq
and carry out the Fourier transformation with respect to the relative
coordinate:
\beq
  \Delta(X, p) = \int d^4(x-y) \, \text{e}^{ip\cdot(x-y)} \Delta(x, y)
.
\label{Wigner_transform:def}
\eeq
Note that the functions $i\Delta^{<,>}(X, p)$, called Wigner functions, are
real, while $i\Delta^t(X,p)$ is the complex conjugate of
$i\Delta^{\bar{t}}(X,p)$. The respective self-energies have the same
properties.

We apply the Wigner transform to the equation of
motion~(\ref{Green_function:EOM:index}). The fact that we are dealing with
two-point functions is reflected in the appearance of an infinite series of
derivatives
\beq
   \diamond \{(1)\} \{(2)\}
 = \frac 12 \left(  \del^{(1)}   \cdot \del_p^{(2)}
                  - \del_p^{(1)} \cdot \del^{(2)}
            \right) \{(1)\} \{(2)\}
\label{diamond:def}
\eeq
with respect to $X$ and $p$:
\beqa
&&
  \left( - p^2 + m^2 -ip\cdot\del
         + \frac 14 \del^2
  \right) \Delta^{<,>}(X, p)
\label{Wigner:EOM}
\\
&&\hphantom{XXX}
  = \text{e}^{-i\diamond}
   \Big(   \{ \Pi^{<,>}(X, p) \} \{ \Delta^A(X, p) \}
          - \{ \Pi^R(X, p) \} \{ \Delta^{<,>}(X, p) \}
   \Big)
.
\nonumber
\eeqa
%

%
%

Now some approximations are necessary.
We assume that the functions of interest have a smooth macroscopic behavior,
more precisely, we assume that the scale on which the functions change with
respect to the average coordinate $X$ is much bigger than the microscopical
scale set by the de Broglie wavelength of the particles. Consequentially, the
mixed derivative $\del\cdot\del_p$ is a small quantity and kept only up to
linear order.
We furthermore assume that the macroscopical scale is also large compared to
the particles' Compton wavelength, so that the second order derivative on the
left hand side of~(\ref{Wigner:EOM}) is negligible as well.
Since the Wigner functions $i\Delta^{<,>}$ are real, we can extract the real
and imaginary part of the equation of motion and find
\beqa
     \left( p^2 - m^2 \right) i\Delta^{<,>}
 &=& -\frac i2
       \left(   \Pi^{<,>} \, \text{Re}\,\Delta^R
              + \text{Re}\,\Pi^R \, \Delta^{<,>}
       \right)
      - \frac 12 \diamond \left( \Pi^< \Delta^> - \Pi^> \Delta^< \right)
,\,\,
\label{EOM:real}
\\
     \left(-p \cdot \del\right) i\Delta^{<,>}
 &=& \frac 12 \left( \Pi^> \Delta^< - \Pi^< \Delta^> \right)
     - \frac i2 \diamond
        \left(   \Pi^{<,>} \, \text{Re}\,\Delta^R
               + \text{Re}\,\Pi^R \, \Delta^{<,>}
        \right)
.
\label{EOM:imaginary}
\eeqa
The real part has the form of a constraint, the imaginary part has the form of
a kinetic equation.

%
%

For a free field we can write down the solutions of these equations in the form
\beqa
  i\Delta_0^<(X, p) &=& 2\pi \delta(p^2-m^2) \text{sgn}(p_0) n_0(X, p)
,
\label{Green_free_<>}
\\
  i\Delta_0^>(X, p) &=& 2\pi \delta(p^2-m^2) \text{sgn}(p_0)
                          \big(1+ n_0(X, p)\big)
\nonumber
.
\eeqa
The solutions for the chronological and antichronological Green functions in
the free case are
\beqa
  i\Delta_0^t(X, p) &=& \frac{i}{p^2-m^2+i\text{sgn}(p_0)\epsilon}
             + 2\pi \delta(p^2-m^2) \text{sgn}(p_0) n_0(X, p)
,
\label{Green_free_ttbar}
\\
  i\Delta_0^{\bar{t}}(X, p) &=& \frac{-i}{p^2-m^2+i\text{sgn}(p_0)\epsilon}
             + 2\pi \delta(p^2-m^2) \text{sgn}(p_0) \big(1 + n_0(X, p)\big)
.
\nonumber
\eeqa
In thermal equilibrium the KMS relation determines $n_0$ to be the
Bose-Einstein distribution, but in a nonequilibrium situation this function is
not known a priori.

%
%

The next simplification is a small coupling expansion. Later this will be used
for a detailed analysis of the collision term, here we need it to get rid of
those terms on the right hand side which are suppressed by both the mixed
derivative $\del\cdot\del_p$ and the coupling constant.
At this point we also demand that our system can be described in terms of
(quasi-)particles. To this end we assume that the right hand side of the
constraint equation~(\ref{EOM:real}) vanishes, turning this equation into a
mass-shell condition.
In the end, equations~(\ref{EOM:real}) and~(\ref{EOM:imaginary}) simplify to
\beqa
     \left( p^2 - m^2 \right) i\Delta^{<,>}
 &=& 0
,
\label{constraint}
\\
     \left(-p \cdot \del\right) i\Delta^{<,>}
 &=& \frac 12 \left( \Pi^> \Delta^< - \Pi^< \Delta^> \right)
,
\label{kinetic}
\eeqa
and we can make an on-shell ansatz for the Wigner functions:
\beqa
  i\Delta^<(X, p) &=& 2\pi \delta(p^2-m^2)
                  \Big[  \theta(p_0)  f_+(X, \vec{p}\,)
                        +\theta(-p_0) \big( 1+f_-(X, -\vec{p}\,) \big)
                  \Big]
,
\label{final:ansatz:Wigner}
\\
  i\Delta^>(X, p) &=& 2\pi \delta(p^2-m^2)
                  \Big[  \theta(p_0)  \big( 1+f_+(X, \vec{p}\,) \big)
                        +\theta(-p_0) f_-(X, -\vec{p}\,)
                  \Big]
.
\nonumber
\eeqa
Spectral sum rules that follow from the basic commutator relations for the
scalar field operators and make a connection between $i\Delta^<$ and
$i\Delta^>$ ensure that this ansatz is consistent.
By comparison with the equilibrium functions and also by inserting the
ansatz~(\ref{final:ansatz:Wigner}) into the expressions for the expectation
values of current
\beq
  j^\mu(X) = 2\int \frac{d^4p}{(2\pi)^4} \, p^\mu \, i\Delta^<(X, p)
\label{current:def}
\eeq
and energy-momentum
\beq
  T^{\mu\nu}(X) = \int \frac{d^4p}{(2\pi)^4} \, p^\mu p^\nu \, i\Delta^<(X, p)
,
\label{energy-momentum:def}
\eeq
we finally can identify $f_+$ and $f_-$ with the phase space densities of
particles and anti-particles, respectively.
In the case of real scalar fields there is an additional relation,
$i\Delta^<(-p)=i\Delta^>(p)$, which leads to $f_+=f_-$.

%
%

The constraint equation is satisfied identically with this ansatz, and all that
is left over is the kinetic equation~(\ref{kinetic}).  We
insert~(\ref{final:ansatz:Wigner}) and integrate over positive frequencies. The
resulting equation reproduces the flow term of a relativistic Boltzmann
equation for particles with phase space density $f_+$:
\beq
  \left( \del_t + \frac{\vec{p}}{\omega}\cdot\vec{\del} \right) f_+(X, \vec{p})
  = \int_0^\infty \frac{dp_0}{\pi} \, C(X, p)
.
\label{kinetic-equation:final}
\eeq
Integration over negative momenta results in a similar equation for the
corresponding antiparticles. The most important part for us is the right hand
side: the collision term so far is
\beq
  C(X, p) = \frac{1}{2}
          \Big(   i\Pi^>(X, p) \, i\Delta^<(X, p)
                - i\Pi^<(X, p) \, i\Delta^>(X, p)
          \Big) 
.
\label{collision-term:def}
\eeq
In the next section we try to re-express this in terms of particle densities
and scattering amplitudes.

%
%
%
%

\section{Collision term}
\label{sec:collision-term}

Now we come to the main part of this paper, the collision term. We make a
perturbative expansion of the self-energies in $C$ and try to bring it to a
form resembling the collision term of a Boltzmann equation like shown in the
introduction.
The collision term is local in our approximation, so we drop the argument $X$
from now on in order to simplify the notation.

\subsection{Self-energy}
\label{subsec:self-energy}

%
%

\begin{figure}[!tb]
\unitlength=1in
\begin{minipage}[t]{2.8in}
\begin{center}
\psfrag{A}[cc]{$\scriptstyle -$}    
\psfrag{B}[cc]{$\scriptstyle -$}    
\psfrag{C}[cc]{$\scriptstyle +$}    
\psfrag{D}[cc]{$\scriptstyle -$}    
\psfrag{E}[cc]{$\scriptstyle +$}    
\psfrag{F}[cc]{$\scriptstyle +$}    
\psfrag{a}[cc]{$\scriptstyle p$}    
\psfrag{b}[cc]{$\scriptstyle p$}    
\includegraphics[width=2.2in,height=1in]{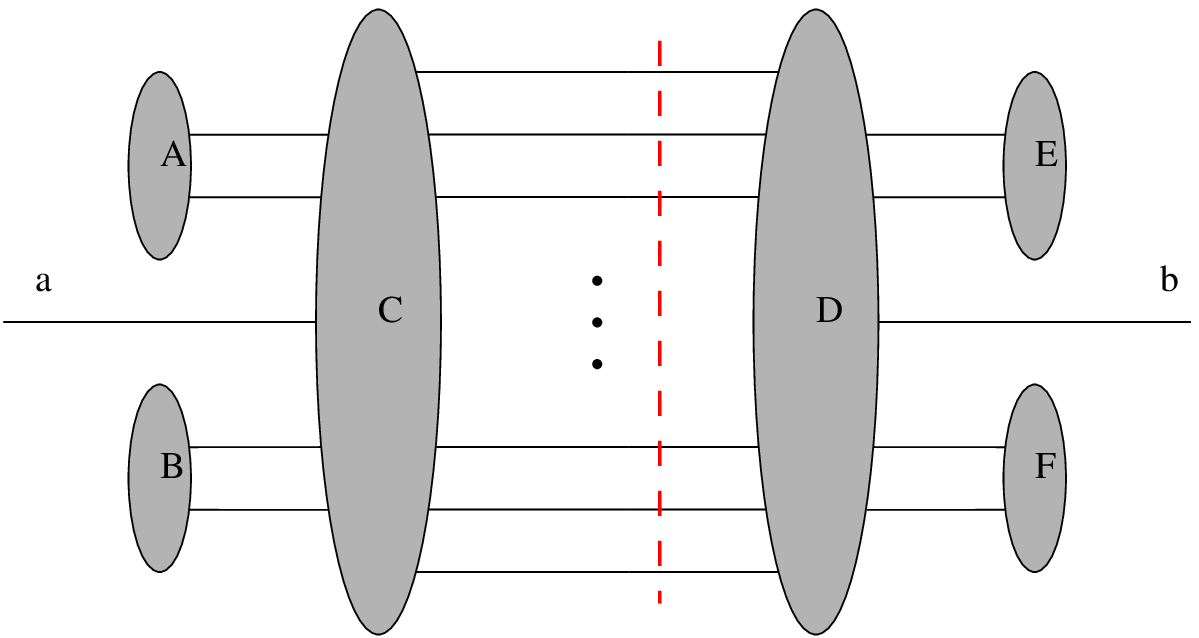}
\end{center}
\caption{\small A graph with only direct connections between
\vp- and \vm-base (type 1).
There is a unique complete cut (dashed line).}
\label{fig:type1}
\end{minipage}
\hfill
\begin{minipage}[t]{2.8in}
\begin{center}
\psfrag{A}[cc]{$\scriptstyle -$}    
\psfrag{B}[cc]{$\scriptstyle -$}    
\psfrag{C}[cc]{$\scriptstyle +$}    
\psfrag{D}[cc]{$\scriptstyle -$}    
\psfrag{E}[cc]{$\scriptstyle +$}    
\psfrag{F}[cc]{$\scriptstyle -$}    
\psfrag{G}[cc]{$\scriptstyle +$}    
\psfrag{H}[cc]{$\scriptstyle +$}    
\psfrag{a}[cc]{$\scriptstyle p$}    
\psfrag{b}[cc]{$\scriptstyle p$}    
\includegraphics[width=2.2in,height=1in]{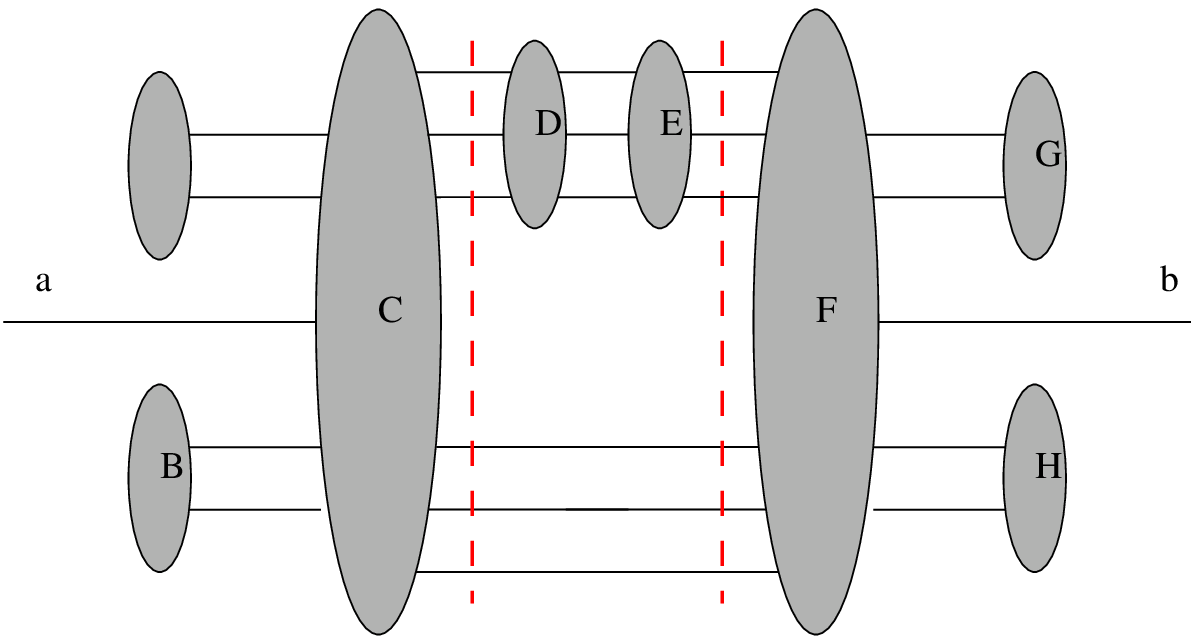}
\end{center}
\caption{\small A graph with indirect connections between
\vp- and \vm-base (type 2).
Two different complete cuts are possible (dashed lines).}
\label{fig:type2}
\end{minipage}
\end{figure}

In a perturbative expansion the self-energy $i\Pi^<(p)=i\Pi^{+-}(p)$ is
expressed as the sum of all amputated one-particle irreducible graphs with
momentum $p$ entering at a \vp-vertex and leaving at a \vm-vertex. We first
classify the graphs in this expansion in the following way:
take any graph and imagine all lines connecting a \vp-vertex with a \vm-vertex
were cut. This would split the graph into a number of connected subgraphs, each
containing only \vp- or \vm-vertices, respectively. We call these subgraphs
{\em clusters}.
Obviously there are at least two clusters, namely one which is connected to the
incoming line, called \vp-base, and one connected to the outgoing line, called
\vm-base.
Now we can distinguish two types of graphs:
\begin{enumerate}
\item
  Graphs of type 1 have only direct connections between the \vp- and the
  \vm-base, i.e. there are no paths leading from the \vp-base to the \vm-base
  via some other clusters (see Fig.~\ref{fig:type1}).  The simplest example for
  this are graphs with only two clusters.
\item
  Graphs of type 2 have connections between the \vp-base and the \vm-base via
  other clusters (see Fig.~\ref{fig:type2}).
\end{enumerate}
There is another way to classify the graphs contributing to $i\Pi^{+-}(p)$:
try to divide a given graph into two parts, one attached to the incoming line,
the other part attached to the outgoing line, by cutting a suitable set of
``$+-$''-lines. Only cut lines which are attached to a \vp-vertex in the first
part and to a \vm-vertex in the second part, and which are effective in
separating the two parts.
This works with every graph, since the incoming line ends in a \vp-vertex,
while the outgoing line starts from a \vm-vertex.  We call this a complete cut.
It is not hard to see that a graph belongs to type 1 if and only if there is
exactly one such complete cut. In this case the cut lines are exactly the ones
connecting the \vp-base with the \vm-base.  Otherwise the graph belongs to type
2.

\subsubsection*{Unique complete cut}

Let us first concentrate on the graphs of type 1, i.e. graphs which have a
unique complete cut. We further classify these diagrams according to the number
of cut lines:
\beq
  i\Pi^{+-}_{(\text{type 1})}(p) = \sum_{n=2}^\infty \, i\Pi^{+-}_n(p)
.
\label{classify-by-n}
\eeq
Now we claim that
\beqa
      -i\Pi^{+-}_n(p)
  &=& \frac{1}{n!}
      \int \frac{d^4k_1}{(2\pi)^4} \, \ldots \, \frac{d^4k_n}{(2\pi)^4}
      (2\pi)^4 \delta^4(p-k_1-\ldots-k_n)
\label{basic-claim}
\\[1ex]
  &&\hphantom{x}
       \times
       {\cal M}^+_{(1)}(p, -k_1, \ldots , -k_n)
       \,
       i\Delta^{+-}_0(k_1) \ldots i\Delta^{+-}_0(k_n)
       \,
       {\cal M}^-_{(1)}(-p, k_1, \ldots, k_n)
.
\nonumber
\eeqa
Symbolically we can write this equation as
\beq
\begin{minipage}{1.3in}
\begin{center}
\psfrag{a}[cc]{$\scriptstyle p$}
\psfrag{b}[cl]{$\scriptstyle p$}
\psfrag{c}[cc]{$\scriptscriptstyle +$}
\psfrag{d}[cc]{$\scriptscriptstyle -$}
\psfrag{e}[cc]{$\scriptscriptstyle (n)$}
\includegraphics[width=1in]{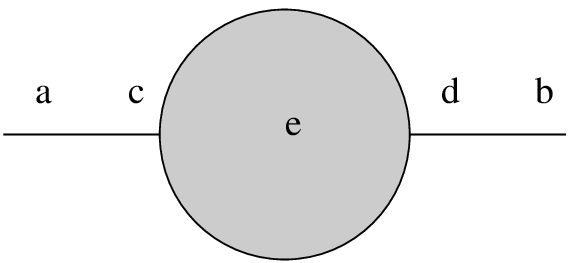}
\end{center}
\end{minipage}
  = \frac{1}{n!} \int \{ dk_i \}
%
\begin{minipage}{1.3in}
\begin{center}
\psfrag{a}[cc]{$\scriptstyle p$}
\psfrag{b}[cl]{$\scriptstyle k_1$}
\psfrag{c}[cl]{$\scriptstyle k_2$}
\psfrag{d}[cl]{$\scriptstyle k_n$}
\psfrag{e}[br]{$\scriptscriptstyle +$}
\psfrag{f}[bl]{$\scriptscriptstyle +$}
\psfrag{g}[bl]{$\scriptscriptstyle +$}
\psfrag{h}[cc]{$\scriptscriptstyle +$}
\includegraphics[width=1in]{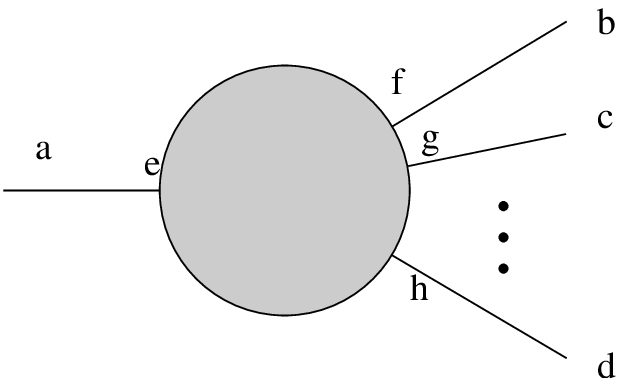}
\end{center}
\end{minipage}
%
\{i\Delta^{+-}_0(k_i)\}
\hphantom{xx}
\begin{minipage}{1.3in}
\begin{center}
\psfrag{a}[cc]{$\scriptstyle p$}
\psfrag{b}[cr]{$\scriptstyle k_1$}
\psfrag{c}[cr]{$\scriptstyle k_2$}
\psfrag{d}[cr]{$\scriptstyle k_n$}
\psfrag{e}[bl]{$\scriptscriptstyle -$}
\psfrag{f}[br]{$\scriptscriptstyle -$}
\psfrag{g}[br]{$\scriptscriptstyle -$}
\psfrag{h}[cc]{$\scriptscriptstyle -$}
\includegraphics[width=1in]{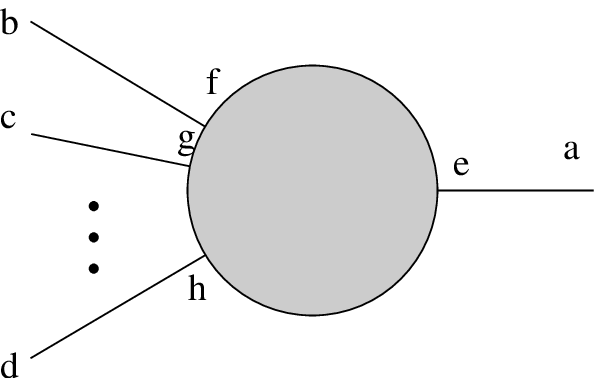}
\end{center}
\end{minipage}
.
\nonumber
\eeq
The infinite sum of diagrams that contribute to the self-energy on the left
hand side factorizes into two other infinite sums which are identified as the
diagrammatic expansions of certain $n$-point functions.
The totally symmetric function ${\cal M}^+_{(1)}$ is the scattering amplitude
belonging to the amputated, connected, out-of-equilibrium $(n+1)$-point
function; positive momenta are entering, and all external momenta are attached
to \vp-vertices.  Basically it is the Fourier transform
\beq
    \Delta_{(1)}^+(q_1, \ldots, q_{n+1})
  = \prod_{j=1}^{n+1}
    \left[
    \int d^4x_j \,
    \text{e}^{iq_jx_j} \,
    \Box_{x_j} 
    \right]
    \Delta_{(1)}^+(x_1, \ldots , x_{n+1})
\eeq
of the out-of-equilibrium $(n+1)$-point function
(we use the same symbol for the function and its Fourier transform)
with time ordered fields
\beq
    \Delta_{(1)}^+(x_1, \ldots, x_{n+1})
  = \left< T \phi(x_1) \ldots \phi(x_{n+1}) \right>_\text{conn}
,
\eeq
however with the additional restrictions that the graphs contributing to this
function must not have any corrections on the $p$-line, and for each $k_i$
there must be a path connecting it to $p$ which only contains \vp-vertices.
The derivative operators $\Box = \del^2 + m^2$ remove the external legs and we
explicitly take out the momentum conservation $\delta$-function,
\beq
    \Delta_{(1)}^+(q_1, \ldots, q_{n+1})
  = (2\pi)^4 \delta^4(q_1 + \ldots + q_{n+1})
    i {\cal M}_{(1)}^+(q_1, \ldots, q_{n+1})
,
\eeq
so that ${\cal M}_{(1)}^+$ has the form of a scattering amplitude.
The functions ${\cal M}^-_{(1)}$ and $\Delta^-_{(1)}$ are defined analogously
with anti-time ordering.

In order to prove~(\ref{basic-claim}), we have to show two things:
first, the set of diagrams appearing in the perturbative expansion on the left
hand side is the same as the set of diagrams one finds on the right hand
side. Here we don't care about statistical factors of diagrams or how often they
appear.
This is done in a second step: it may be possible to obtain the same
diagram by putting together several different contributions from $\Delta^+$ and
$\Delta^-$. The number of such combinations together with the factor $1/n!$ and
the symmetry factors of these contributions must match the symmetry factor of
the resulting diagram on the left hand side.

\subsubsection*{Diagrams}

Any graph $G$ contributing to $\Pi^{+-}_n$ has a unique complete cut consisting
of $n$ lines.
The free propagators $i\Delta^{+-}_0$ associated to the cut lines and the
integrals over their momenta, named $k_1, \ldots, k_n$, are taken out of $G$
and written down explicitly on the right hand side of~(\ref{basic-claim}).
The remainders of $G$ to the ``left'' and to the ``right'' of the cut are
called $G^+$ and $G^-$, respectively.
Obviously $G^+$ is a Feynman diagram with $n+1$ external momenta attached to
\vp-vertices. We took away the $k_i$-lines, so it is amputated.
Since $G$ is one-particle irreducible, there can be no corrections on the
$p$-line, while there may be corrections on the $k_i$-lines.
Because $G$ is of type 1, the cut lines are exactly the ones that connect the
\vp-base to the \vm-base, so all external momenta in $G^+$ are attached to the
\vp-base and thus are connected with each other via paths that only include
\vp-vertices.
This shows that $G^+$ is a Feynman diagram that contributes to
$\Delta_{(1)}^+$, and likewise $G^-$ is a part of $\Delta_{(1)}^-$.

Conversely, take any contributions $G^+$ form $\Delta_{(1)}^+$ and $G^-$ from
$\Delta_{(1)}^-$. Together with the free propagators $i\Delta^{+-}_0$ and the
integrals over their momenta they make up an amputated Feynman diagram $G$ with
a unique complete cut and with momentum $p$ entering at a \vp-vertex and
leaving at a \vm-vertex.
Since there are at least two $k_i$-lines, the only way for $G$ not to be
one-particle irreducible would be to consist of a $G^+$ or a $G^-$ which can be
split into two parts, one connected to the $p$-line and one connected to the
$k_i$-lines, by cutting a single line. But this is impossible, because $G^+$
and $G^-$ don't have corrections on the $p$-line.
So the diagram $G$ contributes to the diagrammatic expansion of $\Pi^{+-}_n$,
which eventually proves that the same diagrams are appearing on both sides
of~(\ref{basic-claim}).

%
%

A comment about the momentum conservation $\delta$-function is in order here.
If one performs as many integrals over internal momenta as possible in the
diagrams $G^+$ and $G^-$, in both cases a $(2\pi)^4\delta^4(p-k_1-\ldots-k_n)$
is left over which is not part of the scattering amplitude.  One appears
explicitly in~(\ref{basic-claim}), the other one reduces to
$(2\pi)^4\delta^4(0)$ and is dropped according to the usual definition of the
self-energy.

\subsubsection*{Numbers}

Let us first have a closer look at the symmetry factor of a diagram $G$
contributing to $\Pi^{+-}_n$.  The symmetry factor of a diagram is the order of
the graph's symmetry group, which contains all permutations of lines and
vertices that do not alter the diagram. $G$ has a unique cut, and accordingly
we can distinguish lines and vertices in the part $G^+$ left of the cut from
lines and vertices in the part $G^-$ to the right of the cut, and they all are
topologically different from the cut lines. This means that there are no
symmetries exchanging lines or vertices in $G^+$ with those in $G^-$, and
neither are there symmetries that exchange cut lines with uncut ones.
Therefore the symmetry group of $G$ is the direct product of three groups: the
symmetry group of $G^+$, the symmetry group of $G^-$, and the symmetry group
$S$ of the cut lines (in the context of $G$, i.e. permutations of cut lines
that do not change $G$).
We call the orders of these groups $s_+$, $s_-$ and $s$, respectively, and so
the symmetry factor of $G$ is $1/(s_+ss_-)$.
In fact, things are a bit more complicated, since in graphs with tadpole-like
structures, i.e. where both ends of some lines are attached to the same vertex,
the above definition of the symmetry group is too narrow. However, such lines
will never be cut, and therefore their symmetry properties are part of the
symmetry groups of $G^+$ or $G^-$ and don't change this discussion.

Given a graph $G$ contributing to $\Pi^{+-}_n$, how many different combinations
of a $G^+ \in \Delta^+$ and a $G^- \in \Delta^-$ produce this $G$?
First of all, the topologies of $G^+$ and $G^-$ are completely determined by
$G$ and its cut. But there are several possibilities to name their external
lines, which determines how they are put together. Some namings result in
topologies different from $G$, so they contribute to a different graph and are
not relevant here.
The rest of the namings corresponds exactly to the inequivalent renamings of
the cut lines $k_i$ of $G$, where inequivalent means that they are not
symmetries of $G$.
So the number of combinations of a $G^+$ and a $G^-$ that produce $G$ is equal
to the number of inequivalent renamings or permutations of the cut lines in
$G$.

The group $P$ of all permutations $\sigma_i$ of cut lines in $G$, i.e.  all
possibilities of renaming them, has the order $n!$
If we build the right cosets of $P$ relative to $S$,
\beq
  S\sigma_1, S\sigma_2, \ldots, S\sigma_{n!}
,
\label{coset:a}
\eeq
then two sets $S\sigma_i$ and $S\sigma_j$ are either identical or
disjoint~\cite{GroupTheory}.  So in fact there are only $f$ different sets
\beq
  S\sigma_{i_1}, \ldots , S\sigma_{i_f}
,
\label{coset:b}
\eeq
each of which contains $s$ elements.
Since all elements within one $S\sigma_i$ belong to equivalent permutations
of cut lines, the number of inequivalent permutations must be equal to $f$.
But the step from~(\ref{coset:a}) to~(\ref{coset:b}) only removes redundant
elements and leaves their total number unchanged: $n!=f \cdot s$. Together with
the symmetry factors of the parts $G^+$ and $G^-$ the complete combinatorial
factor on the right hand side then is
\beq
  \frac{f}{n!}\frac{1}{s_+s_-} = \frac{1}{s_+ss_-}
\eeq
and exactly matches the symmetry factor of $G$.

\subsubsection*{Example}

\begin{figure}[!tb]
\begin{center}
\begin{minipage}[t]{4in}
\psfrag{a}[rt]{$\scriptstyle (a)$}
\psfrag{v1}[cc]{$\scriptstyle +$}
\psfrag{v2}[cc]{$\scriptstyle -$}
\psfrag{v3}[rc]{$\scriptstyle +$}
\psfrag{v4}[lc]{$\scriptstyle -$}
\includegraphics[width=1.2in]{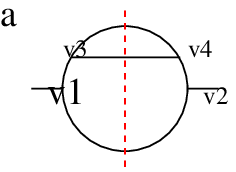}
\hspace*{1in}
\psfrag{a}[rt]{$\scriptstyle (b)$}
\psfrag{v1}[cc]{$\scriptstyle +$}
\psfrag{v2}[cc]{$\scriptstyle -$}
\psfrag{v3}[cc]{$\scriptstyle +$}
\psfrag{v4}[cc]{$\scriptstyle -$}
\includegraphics[width=1.2in]{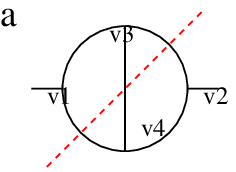}
\end{minipage}
\caption{\small
Two 2-loop contributions to the self-energy of real scalar $\phi^3$-theory with
a unique complete cut comprising $n=3$ lines. The symmetry factors of these
diagrams are $1/2$ and $1$, respectively.  }
\label{fig:phi3}
\end{center}
\end{figure}

As an example we consider two 2-loop contributions to the self-energy of a real
scalar theory with a $\phi^3$-interaction, displayed in Fig.~\ref{fig:phi3}.
There is only one possible topology for the parts $G^+$ and $G^-$, namely a
2-to-2-scattering diagram. There are three possibilities for naming the
external lines of this diagram, so that the relevant contribution is\\[1ex]
\beq
\left(\,
\begin{minipage}{0.6in}
\begin{center}
\psfrag{o1}[cl]{$\scriptstyle k_1$}
\psfrag{o2}[cl]{$\scriptstyle k_2$}
\psfrag{o3}[cl]{$\scriptstyle k_3$}
\includegraphics[width=0.5in, height=0.4in]{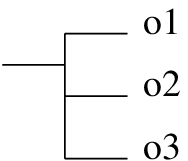}
\end{center}
\end{minipage}
  +
\begin{minipage}{0.6in}
\begin{center}
\psfrag{o1}[cl]{$\scriptstyle k_2$}
\psfrag{o2}[cl]{$\scriptstyle k_1$}
\psfrag{o3}[cl]{$\scriptstyle k_3$}
\includegraphics[width=0.5in, height=0.4in]{Figures/1to3.eps}
\end{center}
\end{minipage}
  +
\begin{minipage}{0.6in}
\begin{center}
\psfrag{o1}[cl]{$\scriptstyle k_3$}
\psfrag{o2}[cl]{$\scriptstyle k_1$}
\psfrag{o3}[cl]{$\scriptstyle k_2$}
\includegraphics[width=0.5in, height=0.4in]{Figures/1to3.eps}
\end{center}
\end{minipage}
\,\right)
\times
\left(\,
\begin{minipage}{0.65in}
\begin{flushright}
\psfrag{i1}[cr]{$\scriptstyle k_1$}
\psfrag{i2}[cr]{$\scriptstyle k_2$}
\psfrag{i3}[cr]{$\scriptstyle k_3$}
\includegraphics[width=0.5in, height=0.4in]{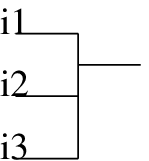}
\end{flushright}
\end{minipage}
  +
\begin{minipage}{0.65in}
\begin{flushright}
\psfrag{i1}[cr]{$\scriptstyle k_2$}
\psfrag{i2}[cr]{$\scriptstyle k_1$}
\psfrag{i3}[cr]{$\scriptstyle k_3$}
\includegraphics[width=0.5in, height=0.4in]{Figures/3to1.eps}
\end{flushright}
\end{minipage}
  +
\begin{minipage}{0.65in}
\begin{flushright}
\psfrag{i1}[cr]{$\scriptstyle k_3$}
\psfrag{i2}[cr]{$\scriptstyle k_1$}
\psfrag{i3}[cr]{$\scriptstyle k_2$}
\includegraphics[width=0.5in, height=0.4in]{Figures/3to1.eps}
\end{flushright}
\end{minipage}
\,\right).
\label{a-times-b}
\eeq\\[1ex]
From the resulting nine graphs, three correspond to the inequivalent ways of
naming the cut lines in diagram Fig.~\ref{fig:phi3}(a), while the remaining six
graphs belong to diagram Fig.~\ref{fig:phi3}(b).

\subsubsection*{No unique cut}

Next we have to deal with those contributions to the self-energy with more than
one complete cut (type 2). In this case we decide to use the cut which makes
the left portion $G^+$ of the diagram as small as possible. This means to cut
all lines that leave the \vp-base, except when they are not efficient in
separating the two parts. In the example of Fig.~\ref{fig:type2} this is the
left cut. This prescription leads to an unambiguously defined cut and we can
repeat all the arguments already used above. The only difference concerns the
nature of the right hand portion of the diagrams:
the contributions to ${\cal M}^-_{(2)}$ must have the property that at least
one of the external $k_i$-lines is connected to the outgoing $p$-line only via
paths which contain at least one \vp-vertex.
Note that ${\cal M}_{(1)}$ and ${\cal M}_{(2)}$ don't have any graphs in
common.

\subsection{Collision term}
\label{subsec:collision-term}

For real fields the $n$-point functions $\Delta^+$ and $\Delta^-$ are related
by complex conjugation:
\beq
   \Delta^-(k_1, \ldots, k_n) = {\Delta^+}^*(-k_1, \ldots, -k_n)
.
\eeq
By referring to the corresponding diagrammatic representation we can check that
this holds for the restricted functions $\Delta^\pm_{(1)}$ and
$\Delta^\pm_{(2)}$, too.
After sending $k_i \to -k_i$ the full collision term can thus be written
\beqa
&&\hphantom{}
    C(p)
  = \frac {1}{2} \sum_{a=1,2} \sum_{n=2}^\infty \frac{1}{n!}
    \int \frac{d^4k_1}{(2\pi)^4} \ldots \frac{d^4k_n}{(2\pi)^4} \,
    (2\pi)^4\delta^4(p+k_1+\ldots+k_n)
\label{Coll:general}
\\&&\hphantom{XX}
    \times \Big\{
                    {\cal M}^+_{(1)}(p, k_1, \ldots, k_n)
                    {{\cal M}^+_{(a)}}^*(p, k_1, \dots, k_n)
                    i\Delta^>_0(k_1) \ldots i\Delta^>_0(k_n) i\Delta^>(p)
\nonumber\\&&\hphantom{XXXX}
                  - {{\cal M}^+_{(1)}}^*(-p, -k_1, \ldots, -k_n)
                    {\cal M}^+_{(a)}(-p, -k_1, \dots, -k_n)
                    i\Delta^<_0(k_1) \ldots i\Delta^<_0(k_n) i\Delta^<(p)
           \Big\}
\nonumber
.
\eeqa
The sum over $a=1,2$ represents the contributions of self-energy diagrams of
type 1 and 2, respectively. We emphasize that this factorization is exact. The
form~(\ref{Green_free_<>}) for the Wigner functions suggests
that~(\ref{Coll:general}) is an expansion in the particle phase space
density. But this is not the case, since the distribution function is also
contained in the scattering matrix. A true expansion in the particle densities
has been done in~\cite{Jeon}, which shares some technical similarities with our
analysis.

In order to proceed towards the standard collision term of a Boltzmann
equation, we still have to overcome three difficulties.
First, a matrix element squared can obviously only be obtained for type 1
diagrams, where $a=1$.
Second, the propagators corresponding to the cut lines are free ones, while the
propagator for the $p$-line is a full one. In the Boltzmann equation, however,
all distribution functions are expected to be of the same type.
And third, if we want to combine the scattering matrices of the last two lines,
they have to be symmetric with respect to inverting all momenta.
The last point is satisfied if we assume that our system is symmetric under
CP. Note that this refers to both the interaction and the initial conditions,
since if the evolution starts with initial particle distributions $f$ that are
not CP-even, then the matrix elements won't be CP-even either, even if the
interaction conserves CP.
The first two points force us to restrict ourselves to the tree
level,
since then the difference between free and full propagators vanishes. In
addition, the scattering diagrams include no internal \vm-vertices: diagrams
with \vm-vertices have several ``$+-$''-lines which cannot all be on-shell in a
tree diagram, and according to~(\ref{Green_free_<>}) these diagrams vanish.

The fact that we have to exclude quantum corrections is not unexpected.  After
all, the Boltzmann equation is a classical equation, so our derivation
definitely has to break down at some level when adding quantum effects. As we
have seen, this breakdown takes place immediately beyond the classical level:
we can recover a standard Boltzmann collision term of the
form~(\ref{Boltzmann}) only by the restriction to classical processes.
Now we can simplify the collision term to
\beqa
    C(p)
  &=&
    \frac{1}{2}
    \sum_{n=2}^\infty \frac{1}{n!}
    \int \frac{d^4k_1}{(2\pi)^4} \ldots \frac{d^4k_n}{(2\pi)^4}
    (2\pi)^4\delta^4(p+k_1+\ldots+k_n)
\label{Coll:type1}
\\&&\hphantom{X}
    \times \big| {\cal M}^+_{(1)}(p, k_1, \ldots, k_n) \big|^2
\nonumber\\&&\hphantom{XXX}
  \times
           \Big(
                  i\Delta^>(k_1) \ldots i\Delta^>(k_n) i\Delta^>(p)
                 -i\Delta^<(k_1) \ldots i\Delta^<(k_n) i\Delta^<(p)
           \Big)
\nonumber
.
\eeqa
The final step is to use the on-shell ansatz~(\ref{final:ansatz:Wigner}) for
the propagators and to perform the integrals over the zero-components of the
momenta. The ansatz for $i\Delta^<(k_i)$ contains two $\delta$-functions, one
that corresponds to positive energies and one that corresponds to negative
energies. In the case of a negative energy, we additionally invert the
corresponding spatial momentum.
This way we obtain contributions with all possible combinations of particles
$k_1$ to $k_n$ either going into or coming out of the scattering.
Since the scattering amplitude is totally symmetric with respect to the order
of its arguments, all particles going in are exchangeable, and so are all the
particles that come out of the scattering. Thus we can group those terms that
have the same number of ingoing and outgoing particles. For $j$ outgoing
particles (besides $p$), there are
\beq
  N(j) = \frac{n!}{j!(n-j)!} = \binom{n}{j}
\eeq
identical terms. The complete collision term is
\beqa
  C(p) &=&
   \frac 12 \frac{\pi}{\omega_p} \delta(p_0-\omega_p) \,
   \sum_{n=2}^\infty \frac{1}{n!} \,
   \int \frac{d^3k_1}{(2\pi)^32\omega_1}
         \ldots
         \frac{d^3k_n}{(2\pi)^32\omega_n} \,\,
   \sum_{j=0}^n \binom{n}{j} 
\label{collision-term:final}
\\ &&\hphantom{XX}
   \times (2\pi)^4\delta^4(p+k_1+\ldots+k_j-k_{j+1}-\ldots -k_n)
\nonumber\\ &&\hphantom{XXXX}
  \times
  \big| {\cal M}^+_{(1)}(p, k_1, \ldots, k_j, -k_{j+1}, \ldots, -k_n)
  \big|^2
\nonumber\\[.1ex]&&\hphantom{XXXXXX}
  \times \Big[
              \big(1+f_1\big) \ldots \big(1+f_j\big)
              f_{j+1}         \ldots f_n
              \big(1+f_p\big)
\nonumber\\&&\hphantom{XXXXXXXXXXX}
      -f_1 \ldots f_j
       \big(1+f_{j+1}\big) \ldots \big(1+f_n\big)
       f_p
  \Big]
,
\nonumber
\eeqa
where $f_i \equiv f(\vec{k}_i)$, $\omega_i=(\vec{k}_i^2+m^2)^{1/2}$, and all
four momenta are on-shell: $k_i=(\omega_i, \vec{k}_i)$.
The collision term consists of two parts, according to the two terms in square
brackets, referred to as gain and loss term, respectively. They describe the
increase or decrease of the density of particles with momentum $p$ in the
plasma due to the scattering.
Typically, several contributions vanish because of kinematical restrictions,
for example the one with all particles going in.  How this works in detail
depends on the type of the interaction and the particle masses, see also the
example in the next section.
Since in the kinetic equation~(\ref{kinetic-equation:final}) we only integrate
over positive $p_0$, terms proportional to the negative energy $p_0=-\omega_p$
were dropped here.

\subsection{Example: $\phi^3$-theory at the 2-loop level}
\label{subsec:example}

\begin{figure}[!tb]
\begin{center}
\begin{minipage}[t]{4in}
\psfrag{i1}[cc]{$\scriptstyle k_2$}
\psfrag{i2}[cc]{$\scriptstyle k_3$}
\psfrag{o1}[cc]{$\scriptstyle p$}
\psfrag{o2}[cc]{$\scriptstyle k_1$}
\psfrag{v1}[cc]{$\scriptstyle +$}
\psfrag{v2}[cc]{$\scriptstyle +$}
\includegraphics[width=1in]{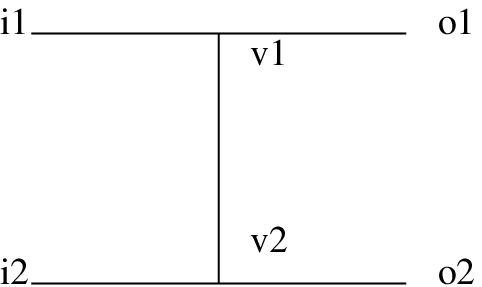}
\hspace*{0.4in}
\psfrag{i1}[cc]{$\scriptstyle k_3$}
\psfrag{i2}[cc]{$\scriptstyle k_2$}
\psfrag{o1}[cc]{$\scriptstyle p$}
\psfrag{o2}[cc]{$\scriptstyle k_1$}
\psfrag{v1}[cc]{$\scriptstyle +$}
\psfrag{v2}[cc]{$\scriptstyle +$}
\includegraphics[width=1in]{Figures/2to2_track.eps}
\hspace*{0.4in}
\psfrag{i1}[cc]{$\scriptstyle k_2$}
\psfrag{i2}[cc]{$\scriptstyle k_3$}
\psfrag{o1}[cc]{$\scriptstyle p$}
\psfrag{o2}[cc]{$\scriptstyle k_1$}
\psfrag{v1}[cc]{$\scriptstyle +$}
\psfrag{v2}[cc]{$\scriptstyle +$}
\includegraphics[width=1in]{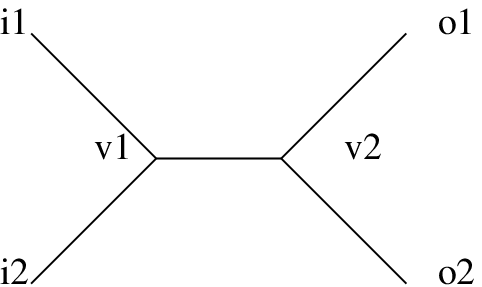}
\end{minipage}
\caption{\small
The tree diagrams contributing to the scattering matrix for four external
particles.}
\label{fig:2to2}
\end{center}
\end{figure}

As an example consider a real, massive scalar theory with a $\lambda\phi^3/3!$
self-interaction. The relevant 2-loop self-energy diagrams are shown in
Fig.~\ref{fig:phi3}.
The cuts in these diagrams comprise three lines, so the corresponding
scattering processes in the Boltzmann collision term will have four external
lines, and at tree level are shown in Fig.~\ref{fig:2to2}. The matrix element
for these diagrams is
\beq
    {\cal M}^+(p, k_1, k_2, k_3)
  = \lambda^2 \,
    \left( \,
           \frac{1}{(p+k_2)^2} + \frac{1}{(p+k_3)^2} + \frac{1}{(p+k_1)^2} \,
    \right)
.
\eeq
The collision term can then be read off from~(\ref{collision-term:final}). In
principle there are 4 contributions, corresponding to 0, 1, 2 or 3 particles
coming out together with $p$, but due to kinematic restrictions only 2-to-2
scattering can occur:
\beqa
  C(p) &=&
   \frac{\pi}{4\omega_p} \delta(p_0-\omega_p) \,
   \int \frac{d^3k_1}{(2\pi)^32\omega_1}
        \frac{d^3k_2}{(2\pi)^32\omega_2}
        \frac{d^3k_3}{(2\pi)^32\omega_3}
\\ &&\hphantom{XX}
   \times (2\pi)^4\delta^4(p+k_1-k_2-k_3) \,
  \big|{\cal M}^+(p, k_1, -k_2, -k_3)\big|^2
\nonumber\\&&\hphantom{XXXXXX}
  \times
  \Big[
      (1+f_p) (1+f_1) f_2 f_3  -  f_p f_1 (1+f_2) (1+f_3)
  \Big]
.
\nonumber
\eeqa
One could have obtained this result directly from the self-energy computed with
the CTP Feynman rules,
\beqa
     -i\Pi^{+-}(p)
 &=& \int \frac{d^4q}{(2\pi)^4} \frac{d^4k}{(2\pi)^4}
\label{self-energy:2-loop}
\\&&\hphantom{XX}
     \Big[
          \frac 12 i\Delta^{-+}(q)     i\Delta^{++}(p+q)
                   i\Delta^{+-}(p+q+k) i\Delta^{-+}(k)
                   i\Delta^{--}(p+q)
\nonumber\\&&\hphantom{XXX}
          +  i\Delta^{++}(p+q) i\Delta^{+-}(p+q+k)
             i\Delta^{--}(q+k) i\Delta^{-+}(q)
             i\Delta^{-+}(k)
     \Big]
,
\nonumber
\eeqa
but already for these comparatively simple diagrams this is not trivial, and it
becomes rather involved for self-energies with more
loops~\cite{Carrington:2004tm}.
%

%
%
%
%

\section{Complex scalar field}
\label{sec:complex}

In the case of a complex scalar field, when there are particles and
antiparticles, we can adopt the general line of argumentation from the real
case, but we have to make adjustments at a few points.

First of all, charge conservation constrains the cut: the net charge flow
across the cut must be equal to the net charge flow through the whole
self-energy diagram. Thus the cut must comprise $2n+1$ lines, $n+1$ leading
from the left to the right, and $n$ leading back from the right to the left.
Then the analog to the basic claim~(\ref{basic-claim}) for complex fields is
\beqa
      -i\Pi^{+-}_n(p)
  &=& \frac{1}{(n+1)!n!}
      \int \frac{d^4k_1}{(2\pi)^4} \, \ldots \, \frac{d^4k_{n+1}}{(2\pi)^4}
      \int \frac{d^4q_1}{(2\pi)^4} \, \ldots \, \frac{d^4q_n}{(2\pi)^4}
\label{basic-claim:complex}
\\[1ex]
  &&\hphantom{X}
  \times
      (2\pi)^4 \delta^4(p + q_1+\ldots+q_n - k_1-\ldots-k_{n+1})
\nonumber\\[1ex]
  &&\hphantom{XX}
  \times
       {\cal M}^+_{(1)}(-k_1, \ldots , -k_{n+1} ; p, q_1, \ldots, q_n)
\nonumber\\[1ex]
  &&\hphantom{XXX}
  \times
       i\Delta^{+-}_0(k_1) \ldots i\Delta^{+-}_0(k_{n+1})
       i\Delta^{-+}_0(q_1) \ldots i\Delta^{-+}_0(q_n)
\nonumber\\[1ex]
  &&\hphantom{XXXX}
  \times
       {\cal M}^-_{(1)}(-p, -q_1, \ldots, -q_n ; k_1, \ldots, k_{n+1})
.
\nonumber
\eeqa
The proof runs almost exactly like in the real case, the only difference is
that we have to distinguish the two sets of cut lines. Lines running from the
left to the right cannot be interchanged with lines running from the right to
the left, because they transport charge in different directions.  This leads to
the factor $1/(n+1)!n!$ that represents one group of $n+1$ lines carrying
charge to the right and another group of $n$ lines carrying charge to the left.

The final result is similar to~(\ref{collision-term:final}), but both gain and
loss term now consist of two parts. In one part the distribution functions
represent a charge coming out of the reaction, where we have all combinations
of these functions being either $(1+f_+)$ for a particle coming out, or $f_-$
for an antiparticle going in. The other part contains distribution functions
that represent a charge going into the reaction, that is all combinations with
either $f_+$ for a particle going in or a $(1+f_-)$ for an antiparticle coming
out.

It is straightforward to generalize to the case of several scalar particle
species in a similar way.  Each species corresponds to a group of lines or
distribution functions, respectively, where members of different groups cannot
be interchanged with each other.

%
%
%
%

\section{Discussion}
\label{sec:discussion}

We start with the equation of motion for the out-of-equilibrium Green function
for a real scalar field.
Using gradient expansion, on-shell approximation and small coupling expansion
the left hand side leads to the flow term of a Boltzmann equation in a well
known way.
We make a perturbative expansion of the self-energies on the right hand side
and give a diagrammatic proof that
the relevant self-energies have a factorization property, and that
for a certain type of contributions to the self-energy the right hand side can
be rewritten as the matrix element squared of scattering processes times the
distribution functions of the particles involved in the reactions. This works
for those self-energy contributions that correspond to tree level scattering
diagrams. The scattering processes obtained include any number of external
particles and thus correspond to self-energy diagrams with any number of loops.

If effects beyond the classical level are required, it is in general not
correct to simply compute the scattering probabilities including quantum
corrections and plug them into a standard Boltzmann collision term.
In specific situations it is possible to fit quantum corrections into the
picture, however at some cost.
In~\cite{Isert:2001kq} for example, a model of scalar quarks and gluons was
examined and a standard Boltzmann collision term was constructed from the
self-energy up to two loops.
In order to be able to do this, the scattering matrices were not computed
simply by following the CTP rules, but for some diagrams Feynman propagators
had to be used for some internal lines.
Furthermore, kinematical arguments had to be used to get rid of some diagrams
that didn't match into the picture.
The approach presented there required an explicit study of each diagram and can
not be generalized to arbitrary diagrams or theories.

The simple structure~(\ref{Boltzmann}) cannot be obtained beyond the classical
level for two reasons.
First, there are the self-energy diagrams that allow several complete
cuts. This is precisely the type of diagrams that causes problems in the
extension of the vacuum Cutkosky rules to finite
temperature~\cite{Kobes,Gelis:1997zv}.
Second, there is the difference between the free propagators for the cut
$k_i$-lines and the full propagator for the $p$-line.
Since we know that the scattering diagrams we obtain must not contain
corrections on the $p$-lines but do have corrections on the $k_i$-lines, it
seems tempting to collect these corrections and attribute them to the free
propagators, thus promoting them to full ones.
In effect this means that we would have to do the perturbative analysis of the
self-energy in terms of resummed and thus full propagators.  As a consequence,
all diagrams contributing to the self-energy must not have corrections to
already full lines.
Then the available self-energy diagrams are not sufficient to be rewritten as
the square of a matrix element, however: certain diagrams are missing. For
instance, the 2-loop diagram in Fig.~\ref{fig:phi3}(b) provides only the
interference terms in~(\ref{a-times-b}), the squares of the individual
amplitudes can only come from Fig.~\ref{fig:phi3}(a), which has a forbidden
correction.

This shows that there is no consistent and systematic way to obtain a standard
Boltzmann collision term beyond the tree level.  At the tree level, a standard
Boltzmann collision term is found that includes scattering processes with any
number of external particles.

%
%
%
%

\end{document}